\documentclass[sn-mathphys,Numbered]{sn-jnl}

\usepackage{graphicx}%
\usepackage{multirow}%
\usepackage{amsmath,amssymb,amsfonts}%
\usepackage{amsthm}%
\usepackage{mathrsfs}%
\usepackage[title]{appendix}%
\usepackage{xcolor}%
\usepackage{textcomp}%
\usepackage{manyfoot}%
\usepackage{booktabs}%
\usepackage{algorithm}%
\usepackage{algorithmicx}%
\usepackage{algpseudocode}%
\usepackage{listings}%

\theoremstyle{thmstyleone}%
\theoremstyle{thmstyletwo}%

\theoremstyle{thmstylethree}%

\begin{document}

\title[Article Title]{A High-performance Real-time Container File Monitoring Approach Based on Virtual Machine Introspection}


\author[1]{\fnm{Kai} \sur{Tan}}\email{20b903076@stu.hit.edu.cn}

\author*[1]{\fnm{Dongyang} \sur{Zhan}}\email{zhandy@hit.edu.cn}

\author[1]{\fnm{Lin} \sur{Ye}}\email{hityelin@hit.edu.cn}

\author[1]{\fnm{Hongli} \sur{Zhang}}\email{zhanghongli@hit.edu.cn}

\author[1]{\fnm{Binxing} \sur{Fang}}\email{fangbx@hit.edu.cn}

\author[2]{\fnm{Zhihong} \sur{Tian}}\email{tianzhihong@gzhu.edu.cn}

\affil*[1]{\orgdiv{School of Cyberspace Science}, \orgname{Harbin Institute of Technology}, \orgaddress{ \city{Harbin}, \postcode{150001},  \country{China}}}

\affil*[2]{\orgname{Guangzhou University}, \orgaddress{ \city{Guangzhou}, \postcode{510000},  \country{China}}}


\abstract{As cloud computing continues to advance and become an integral part of modern IT infrastructure, container security has emerged as a critical factor in ensuring the smooth operation of cloud-native applications. An attacker can attack the service in the container or even perform the container escape attack by tampering with the files. Monitoring container files is important for APT detection and cyberspace security. Existing file monitoring methods are usually based on host operating system or virtual machine introspection to protect file security in real time. The methods based on the host operating system usually monitor file operations in the host operating system. However, when the container escapes to the host, the host operating system will no longer be secure, so these methods face the problem of weak security. The methods based on virtual machine introspection usually monitor the file operations in the virtual machine in real time in the virtual machine monitor layer. Due to the strong isolation ability of the hypervisor, compared with the file monitoring based on the host operating system, the monitoring program in the virtual machine monitor layer is more secure. However, virtual machine introspection technology usually introduces high real-time monitoring overhead. Aiming at the problems of low security and high overload introduced in existing container file monitoring, a high-performance container file monitoring method based on virtual machine introspection is proposed. Based on the container-in-VM architecture, the virtual machine introspection technology is used to monitor the in-VM container files. Based on the isolation capability of the hypervisor, the security problems of security monitoring introduced by container escape attacks can be addressed. To reduce the monitoring overload introduced by file monitoring based on virtual machine introspection in container scenarios, a high-performance real-time file monitoring method based on memory monitoring is proposed. After analyzing the container file system, independent memory areas are initialized to store the memory cache of the monitored container files. Then, the virtual machine introspection technology is used to monitor the target memory areas, so as to capture and analyze the access operations to the target files. Since the monitored files are stored in separate memory areas, monitoring will not affect the read and write performance of other files in the container. The experimental results show that the proposed approach can effectively monitor the container files and introduce an acceptable monitoring overload.}




\maketitle

\section{Introduction}\label{sec1}

In recent years, cloud computing has become an integral part of the modern IT landscape, enabling organizations to deploy, manage, and scale applications efficiently. Cloud computing platforms, such as Amazon Web Services (AWS), Azure, and Google Cloud, have transformed the way applications are built, deployed, and managed. One of the key innovations driving this transformation is the adoption of container technology, which has emerged as a prominent virtualization infrastructure in cloud-native environments due to its lightweight nature and high elasticity \cite{bib1}.

Containers provide numerous benefits, such as rapid deployment, improved resource utilization, and easier application management. Major cloud computing providers have embraced container technology and offer dedicated container services, such as AWS's Elastic Container Service (ECS) and Azure's Azure Container Instances (ACI). However, the rapid growth of container technology has also exposed various security challenges, necessitating the protection of critical files in real-time.

Researchers have discovered vulnerabilities like CVE-2019-5736, which can be exploited by attackers to rewrite the runc binary of the container in the host and execute container escape attacks. In the context of cloud computing, where multiple users and organizations share resources, the security implications of such vulnerabilities are magnified. 
An attacker can not only threaten the security of the host machine but also compromise the isolation between other containers and virtual machines, amplifying the potential security risks. In highly concurrent and integrated cloud platforms, container escape can lead to cross-container attacks, data breaches, or even the collapse of the entire cloud platform.
As a result, effective monitoring and protection of container files are crucial for advanced persistent threat (APT) detection and cyberspace security in cloud computing environments \cite{bib1}.

To protect file security, researchers usually adopt two methods: polling monitoring \cite{bib2} and real-time monitoring \cite{bib3,bib4}. The monitoring approaches are usually implemented as programs or kernel modules in the host to check the security of files or monitor file access operations in real-time. Specifically, polling file monitoring usually periodically checks the integrity of files or compares files with virus databases to find malicious files \cite{bib2}. 
This approach has inherent limitations, as it cannot provide real-time detection or prevent breaches as they occur. Additionally, polling often results in significant performance overhead due to the need for frequent checks, which can negatively impact the efficiency of containerized applications.
In contrast, real-time file monitoring can capture file operations in real-time based on the operating system and analyze its security, which is more suitable for protecting critical files (such as configuration files, and user privacy data) \cite{bib3,bib4}. However, these security monitoring methods running on the host are facing the threat of container escape attacks. The container shares the operating system kernel with the host, resulting in weak isolation capability. Attackers in the container can exploit the vulnerabilities in the operating system to achieve container escape attacks, thereby obtaining the root privilege of the host and threatening the security of the programs in the host \cite{bib5,bib6,bib7,bib8}.

In order to enhance the isolation capability of containers, cloud computing service providers such as OpenStack and Google have proposed new container architectures based on virtual machines, such as Kata Containers and gViosr \cite{bib9}. Under these container architectures, the container runs in a lightweight virtual machine, which provides better isolation capabilities. The virtual machine minimizes the operating system kernel, accelerates the startup speed, and optimizes the resource consumption, which can achieve a trade-off between performance and security, so it is widely used in cloud computing such as Alibaba Cloud and Baidu Cloud.

Under the above container architectures, file monitoring technology based on virtual machine introspection (VMI) \cite{bib10} can effectively improve the security of file monitoring. Based on the virtual machine introspection architecture, the monitoring program can run in the hypervisor, and the monitoring target is in the virtual machine. Due to the strong isolation capability of the hypervisor, 
even if the attacker escapes the container and controls the virtual machine, he still cannot threaten the monitoring programs running in the hypervisor. There are two types of file monitoring based on VMI: polling monitoring and real-time monitoring. Compared with polling monitoring, real-time monitoring can protect the integrity of important files and has wider application scenarios. The real-time file monitoring method based on virtual machine introspection usually sets the watch points at some file access operations in the virtual machine kernel. Whenever a file is accessed, the monitoring program will be triggered to check whether the accessed file is the target monitoring file, and the access operations will be analyzed for security. Due to the high frequency of file access in the virtual machine, this monitoring method based on intercepting file operations will frequently interrupt the normal execution of the virtual machine, introducing a high monitoring overhead, and cannot meet the performance requirements of containers.

To make the real-time file monitoring based on virtual machine introspection meet the needs of container scenarios, a high-performance real-time file monitoring method is proposed based on memory monitoring. Firstly, a container file system analysis method is proposed based on virtual machine introspection, which can automatically analyze and identify the container files in the virtual machine. After that, a separate memory area is established in the virtual machine to store the cache of the target file, and the memory area is monitored. All access to the target file by the virtual machine will trigger the monitoring, while access to other files will not trigger the monitoring, which can efficiently reduce the overhead introduced by real-time monitoring. When monitoring is triggered, the security of file access can be analyzed out of the virtual machine, and abnormal file access will be blocked in real-time.

In order to perform cache-based file monitoring, an automatic file metadata cache creation approach is proposed based on the cooperation between virtual machine and hypervisor. Before the container starts, the virtual machine operating system automatically creates a separate memory area and establishes memory caches of target files, and then the target memory area is monitored based on VMI outside the virtual machine. The automatic file metadata cache creation solves the problem that VMI is difficult to control the execution state of virtual machine due to the semantic gap problem, and improves the speed of cache creation. Therefore, it can meet the performance requirements of containers. 
In addition, VMI allows for monitoring the in-memory state of the virtual machine, enabling more accurate detection of file access and changes within containers without the need for intrusive and resource-heavy polling.

The contributions of this paper are as follows:
\begin{itemize}
   
 \item A high-performance container file monitoring method based on VMI-based memory monitoring is proposed. By monitoring the cache corresponding to the container file in the virtual machine memory, the overhead introduced by VMI-based real-time file monitoring is reduced.

 \item  An automatic file metadata cache creation approach based on the cooperation between virtual machine and hypervisor is proposed, which uses the virtual machine operating system to automatically create a separate memory area and establish the file metadata caches, solving the problems of the semantic gap and the slow cache creation speed faced in VMI. 
 
 \item  The effectiveness, performance and overhead are evaluated after the prototype system is implemented. The experimental results show that the proposed method can effectively monitor the container file and the monitoring overhead is the lowest among the existing methods. In addition, the monitoring program can resist container escape attacks due to the strong isolation ability of the hypervisor.

\end{itemize}

The rest of this paper is organized as follows. Section \ref{sec2} introduces the related works. The system design of our approach is described in section \ref{sec3}. Section \ref{sec4} describes the implementation of the system, and section \ref{sec5} evaluate and discuss our system. Finally, we conclude our work in section \ref{sec6}.

\section{Background}

In the modern landscape of cloud computing, Virtual Machine Monitors (VMMs) such as KVM and Xen play pivotal roles. KVM supports a full virtualization approach, providing a traditional virtualized environment where each guest OS believes it is running on its own hardware. Conversely, Xen has evolved to offer a hybrid model termed Paravirtual Hardware Virtual Machine (PVHVM). This model enhances the performance of virtualized environments by integrating paravirtualized drivers, which reduce the overhead from emulating disk and network operations, thereby boosting I/O efficiency.

Containers serve as a streamlined alternative to virtual machines, encapsulating applications and their dependencies within a single deployable unit. This allows for swift, consistent, and scalable application deployment across various computing environments, utilizing core features of the host operating system. However, container technology often faces challenges with isolation, making robust security measures essential. Leading containerization technologies include LXC, Docker, and Podman. LXC, an early container technology, leverages Linux kernel features like cgroups and namespace isolation to provide secure environments for applications. Docker, known for its ease of use and efficiency, automates application deployment within containers, employing an image-based management approach. Podman, developed by RedHat, offers a Docker-compatible platform that doesn't require daemon processes, enhancing security and simplicity in operations.

In light of the increasing demand for improved isolation in container environments, numerous cloud service providers, such as OpenStack, have turned to virtual machine-based container architectures like Kata Containers. These frameworks utilize lightweight virtual machines to house containers, enhancing isolation through a minimized operating system kernel. This design not only boosts startup times but also enhances resource efficiency, effectively balancing performance with security. This methodology is now widely implemented across leading cloud platforms, including Alibaba Cloud and Baidu Cloud.

\section{Related works}\label{sec2}

File security is an important part of system security, researchers have proposed various methods to protect file security. Under the traditional host architecture, file protection tools are generally used as kernel modules or user-mode programs to protect file security in a polling or real-time method \cite{bib2,bib3,bib4,bib5,bib6}. For example, security tools such as Kaspersky Anti-Virus can compare files with virus databases and find virus files, and ICAR \cite{bib2} can detect the integrity of files by polling monitoring and restore tampered files. \cite{bib3,bib4} run inside the operating system to capture and analyze the security of file access in real-time by intercepting file system calls. Since these security tools and malicious programs are running on the same operating system, attackers may find and even break these security tools.

With the development of virtual machine technology and cloud computing, more and more computing tasks are running in virtual machines, and a new security monitoring technology represented by VMI has been proposed. Under this architecture, the monitoring object runs in the virtual machine, and the monitoring program runs in the virtual machine management layer with higher privileges and is strictly isolated from the monitoring object, so the monitoring program is more secure. File monitoring based on VMI is divided into two methods: polling monitoring and real-time monitoring.

The polling file detection method based on VMI usually obtains the content or attributes of the file out of the virtual machine, and compares them with the malicious file library or file backup to find malicious files or tampered files. Specifically, \cite{bib11} extracts the files in the virtual machine based on VMI and perform security analysis. Polling detection generally does not affect the normal operation of the monitored virtual machine, and is usually used to detect the security of large-scale files in cloud computing. Its representative products include VMware vShield Endpoint \cite{bib12} and so on. However, critical files related to system operation (such as container runc files) need to be protected in real time and cannot be protected by polling monitoring.

The real-time file protection based on VMI usually captures the file operation behavior of the virtual machine, and then analyzes its security. Specifically, \cite{bib13} and Gemini \cite{bib14} set up monitoring points at the file-related system calls in the virtual machine, so as to capture and analyze the security of all file access in the virtual machine. \cite{bib15,bib16} capture the file access at the file block level and analyze its security by modifying the back-end driver of the virtual machine. However, these file monitoring methods are based on monitoring all file operations in the virtual machine, and all file operations in the virtual machine will trigger monitoring and introduce high monitoring overhead. Different from these methods, CFWatcher \cite{bib17,bib18} reduces the monitoring overhead by monitoring the cache of the target file, however, CFWatcher faces several problems: 1) Monitoring the target file metadata cache will affect other files in the adjacent memory, and the overall file performance of the virtual machine will decrease; 2) The file metadata cache is created slowly, so it cannot meet the fast startup requirements of the container scenario; 3) The container file system cannot be analyzed.

With the rapid development of container technology, container security issues have also attracted much attention. Current research mainly focuses on container image security \cite{bib19,bib20,bib21}, container isolation \cite{bib22,bib23}, and behavioral security \cite{bib24,bib25}, but lack of research on container file security. Therefore, a real-time monitoring method for container files is proposed based on VMI, so that real-time file monitoring can meet the needs of container scenarios.

\section{System design}\label{sec3}
\begin{figure}[h]%
\centering
\includegraphics[width=0.65\textwidth]{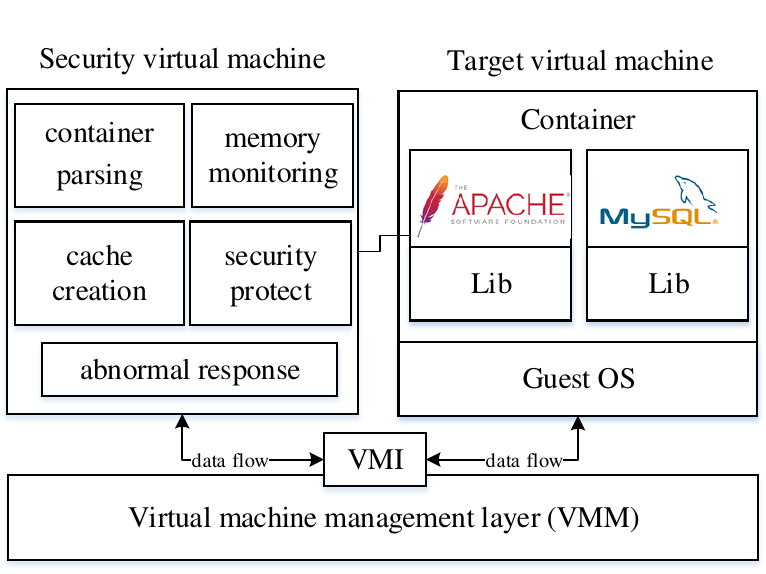}
\caption{System architecture}\label{fig1}
\end{figure}

A high-performance container file monitoring approach is proposed based on VMI, and its system architecture is shown in Fig.\ref{fig1}. Xen \cite{bib26} is used in our system to create virtual machines, which manages access to memory and hardware resources through a combination of its hypervisor and a specially privileged VM called Domain 0. The virtual machine management layer can virtualize various hardware to support the execution of the virtual machines, so it can access the underlying binary state of the target virtual machine and have the highest privilege. The monitored containers are running inside the target virtual machine. The analysis processing program runs in the secure virtual machine (Domain 0), and analyzes the containers in the target virtual machine through the VMI module. The VMI module can transmit the underlying binary running state of the target virtual machine to the secure virtual machine, and can also change the running state of the target virtual machine according to the instructions of the secure virtual machine, thereby realizing abnormal response. The Security monitoring system mainly includes 5 modules: 1) the container parsing module can parse the file system and running information of the containers in the hypervisor; 2) the cache creation module can cooperate with the target virtual machine to create separate memory area and store the cache of monitored files in the virtual machine according to the security requirements before the container is started; 3) the memory monitoring module can capture all operations on the target file in the virtual machine and analyze the security of file access; 4) the abnormal response module can block the malicious file access in real time when abnormal file access is found; 5) the security protection module can protect the security of the monitoring related code and data.

\subsection{Container parsing}\label{subsec3.1}

The establishment of the container file metadata cache is completed by the cooperation of the operating system, so it is necessary to parse and map the container file to the host file first. To this end, an automated container file system parsing approach is proposed.

At present, overlay File System  \cite{bib27} is widely used in containers. It is an implementation of union file system. It can merge file directories stored in different locations and mount them in the same directory. The container file system is usually composed of multiple layers of images, and the lower layer of images is read-only. When a new container is created, an overlay image will be generated. Changes to the files in the underlying image will be saved in the overlay image. Specifically, when a container adds, deletes, and modifies files, a copy will be generated in the upper image layer. The file images of each layer of the container are mounted in the same directory, and each layer has an image ID. Therefore, the files in the container can be parsed by establishing the corresponding relationship between the image ID and the container ID.

In order to analyze the security of the captured file access, the container processes need to be identified and analyzed outside the virtual machine. At present, the container adopts the operating system virtualization technology to isolate the resource views of each container through the namespace technology in the operating system. The Linux operating system has six namespaces. The PID namespace enables the container to have an independent PID sequence number space; The UTS namespace can isolate the container and the host name of the operating system; The Mount namespace enables the container to have an independent file view; The IPC namespace and network namespace are respectively used to isolate inter process messages and network stacks; The User namespace is used to represent the permissions of the namespace. Processes with the same namespace can share the same resource view, so processes in the same container have the same namespace. Based on the characteristics of container processes, we first use VMI to extract the process list in the virtual machine. All processes in Linux are connected by a doubly linked list, and all processes can be identified by parsing the memory of the virtual machine. After that, according to the relationship between the process and the namespace, the namespace corresponding to each process is parsed so as to automatically identify the container in the virtual machine. In addition, by reconstructing the process structure, information such as the process name and user of the container process can be further extracted for subsequent security analysis.

\subsection{File monitoring based on memory monitoring}\label{subsec3.2}
\begin{figure}[h]%
\centering
\includegraphics[width=0.5\textwidth]{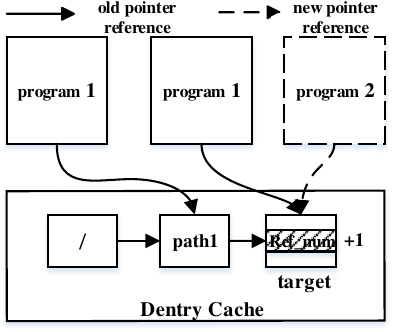}
\caption{The impact of file access on dentry}\label{fig2}
\end{figure}

To discover the file events only related to the target file, \cite{bib17,bib18} analyzed the file management mechanism of the operating system. In the operating system, files usually contain metadata (or attribute information) and file content. When files are accessed, the operating system creates caches to improve I/O device access performance. 
We make full use of the existing dentry cache (dcache) mechanism in the Linux kernel. Since the dcache itself is implemented based on hash tables, our scheme quickly locates the corresponding dentry object by hashing the file pathname, thereby improving access and monitoring efficiency. In this way, our cache design maintains consistency with kernel mechanisms while achieving high efficiency.
Specifically, when accessing a file or directory, the operating system will generate a dentry object in memory, which will record the metadata (or attribute data) of the file, such as file name, creation time, access rights, etc. As shown in Fig.\ref{fig2}, when accessing "/path1/target", the operating system will create dentry objects corresponding to "/", "path1", and "target" according to the level of the file path. And when a new program accesses the file, the operating system will refer to the object that already exists. When multiple programs reference the same object, the reference count is used to record the number of references. Therefore, new file accesses increase the reference count of dentry. Similarly, when a program closes a file, the operating system reduces the corresponding reference count. If the reference count is reduced to zero, the file is not accessed by any program, and the corresponding dentry may be released by the operating system.

Based on the above file access mechanism, all accesses to the file can be captured by monitoring changes in the dentry structure of the target file.  The monitoring program first obtains the memory address of the dentry object of the monitored file. 
Specifically, we use Virtual Machine Introspection (VMI) to obtain the memory address of the dentry object associated with a target file by navigating the kernel data structures of the guest OS from outside the virtual machine. Starting with the task\_struct of the target process, which holds essential process information, we access its files\_struct pointer, containing details of all open file descriptors. Within files\_struct, the fd\_array holds pointers to file structures for each open file descriptor. By iterating through this array, we locate each file structure and its associated dentry object, which represents the directory entry for the file and is essential for monitoring purposes. 

The monitoring program then locates the reference count field in the dentry structure based on the starting address and memory distribution of the dentry structure. Finally, based on VMI, the memory page is read and written to the monitor. When the monitoring is triggered, it first determines whether the target address of the memory write is the reference count of the monitored dentry. If the reference count increases, it means that a new program is accessing the file.

Therefore, the security of file access can be analyzed in real-time based on the security policy. For example, in the basic security policy, the runc file in the virtual machine should be configured not to be modified by any process. In addition to the basic security configuration, users can protect their files with other security policies. For example, the user can set the "/www/target.html" file in container A to be readable only by httpd. When the file is accessed, the container ownership and specific information of the current file access process can be determined based on the analysis result of the container process, and then it can be determined whether the process is httpd in the container. Since it is in the process of the file access system call, when an abnormality is detected, the system can respond to the abnormality in real-time.

At present, the minimum granularity of the memory monitoring of the VMI is the memory page, while the proposed method only needs to monitor one field in the target file dentry, so other areas in the memory page where this field is located are also within the scope of memory monitoring. Reading or writing other areas in these memory pages will also trigger monitoring, and the monitoring trigger conditions are expanded. Therefore, an independent cache area creation method is proposed in Section 3.3 to solve the problem of trigger conditions expansion.

\subsection{Automatic file metadata cache creation}\label{subsec3.3}

The proposed method captures access to the target file by monitoring the cache object of the file. Therefore, the dentry object of the monitored file must be located in the memory of the virtual machine. \cite{bib17,bib18} propose a file metadata cache creation method based on VMI. The core idea is to open the target file by injecting an OPEN system call into the virtual machine. File access is implemented based on the OPEN system call in the operating system, which creates the dentry object of the accessed file in the kernel. Therefore, by injecting the open system call, the virtual machine operating system will automatically create a cache of files. Specifically, the monitoring program waits for a program with root privilege in the virtual machine to invoke any system call and replaces the call number and parameters of the system call, so that the system call is converted into an OPEN system call and the target file is opened, thereby establishing a file metadata cache in the kernel. When multiple files need to be monitored, multiple system calls need to be injected continuously. However, this method is not suitable for the container environment because of its slow execution speed. Firstly, whether there are enough programs with root privilege in the virtual machine to invoke system calls is uncontrollable; Secondly, after the system call is injected, the original system call needs to be executed again to ensure the normal execution of the program, which will also consume a lot of time; In addition, the waiting time is usually long. The start time of the container is very important to the performance of the container. Because these tasks need to be completed before the container is started and consume a lot of time, this method is not suitable for the container scenario.

\begin{figure}[h]%
\centering
\includegraphics[width=0.5\textwidth]{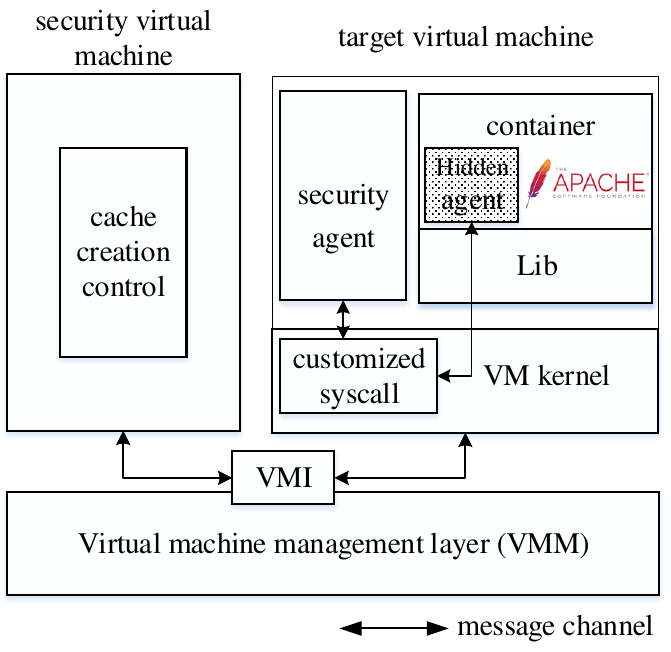}
\caption{The system architecture of file metadata cache creation}\label{fig3}
\end{figure}

To solve these problems, a file metadata cache creation method is proposed based on the cooperation of virtual machine and hypervisor. The system architecture of this method is shown in Fig.\ref{fig3}. The method includes two parts: 1) a security agent running in user mode and a customized system call running in the kernel; 2) a VMI module running outside the virtual machine to control and protect the modules in the virtual machine.

The security agent program first run after the virtual machine is started, which will open the files to be monitored in turn according to the content of the configuration file.
It is important to note that the configuration file is only used before the virtual machine starts. Once the system is running, the configuration file is no longer needed and therefore does not require continuous protection during runtime.
After all files are opened, the container will be opened, and a cache creation completion signal will be sent to the control program running in the virtual machine management layer. For files to be protected in the virtual machine, the security agent is opened in read-only mode. Since the security agent runs with root privileges, it can access all privileged files. For the key files in the container that need to be monitored, the security agent will create a hidden process in the container, which will open the corresponding file in the container file system. The hidden process runs in the container, so it does not need to convert the monitored file path, which has higher efficiency and better versatility.

It is worth noting that our monitoring mechanism reduces the number of dentry objects in memory by limiting the number of monitored files, focusing only on those critical for security purposes, thereby minimizing memory usage. At the same time, the system dynamically manages memory, allowing the operating system to reclaim dentry objects associated with less critical files once memory usage reaches a predefined threshold, achieving a balance between system performance and resource utilization.

In order to make the agent program running in the container hidden to the container process, we adopt the idea of the kernel rootkit to hide the agent program in the container. Kernel rootkits are usually used to hide processes in the operating system, by injecting process hiding code into the system call to remove the process that needs to be hidden from the system call return list. Based on this approach, some key kernel functions are modified in the virtual machine operating system and the process hiding code is injected into several system calls (e.g., getdents). Unlike common rootkits, the agent program in the container only needs to be hidden to the container process and should be visible to the security agent outside the container. Therefore, the injected code needs to identify the container process and the host process. The pid\_ns fields of different container processes will point to different PID namespaces, while the host process will point to the initial namespace, so the process belongs to the container can be identified by identifying the pid\_ns pointer of the container process. When the process in the request process list belongs to the container, the code injected in the kernel will delete the agent program from the process list.

In order to make the dentry structure of the monitored file in an independent memory space, an independent file access system call is added to the virtual machine. The security agent/hidden agent can make the monitored file metadata cache in an independent memory through this system call. By default, the operating system will find the available space in the memory according to the size of the kernel object to be allocated. This will make the target object and other objects in the same memory page. However, the system call added by the cache creation module adopts a different memory allocation method. It will first allocate a new memory page, and then place the memory object in the page. Based on this memory placement method, the monitored object and other files can be located in different memory pages. The read and write monitoring of these pages will not affect the virtual machine's access to other memory pages, thereby reducing the overhead introduced by monitoring. In order to improve memory utilization, the proposed method does not place the dentry objects of each monitored file in a separate memory page, but places them in adjacent memory spaces. Therefore, the same memory page may contain multiple monitored dentry objects, thereby improving memory utilization.

The security agent reads security configuration policies from the virtual machine before the container starts, and these policies are written to the virtual machine image through image editing before the virtual machine starts. Through image editing, the virtual machine can already have all the security policies before starting, thereby shortening the startup time of the container in the virtual machine. After the target file metadata cache is established, the security agent needs to transmit the start monitoring signal and the memory address to be monitored in the virtual machine to the memory monitoring module. In addition, when the files generated during the running of the container are monitored, the security configuration needs to be delivered to the security agent or the hidden agent according to the requirements.

To implement an information transmission mechanism based on Virtual Machine Introspection (VMI), this paper introduces a custom system call named Monitor\_Syscall into the virtual machine kernel. The primary purpose of this system call is to serve as a carrier for transmitting data required by the security agent and the hidden agent. Specifically, Monitor\_Syscall is added to the kernel, which includes a series of NOP instructions and immediately returns to user mode upon execution, ensuring that it does not significantly impact the normal system call flow. After the virtual machine boots, The VMI module in the virtual machine management layer will replace the first instruction of this system call with an interrupt instruction INT3, triggering a VMExit event.Because directly modifying the kernel to add a new interrupt-based system call compromises system stability.

When Monitor\_Syscall is invoked by security agent, the VMI module captures the VMExit event and reads the parameters of the system call, which contain the data transmitted by the security or hidden agent. This data is used to convey monitoring commands or configuration information. The VMI module can modify the virtual machine's memory to pass information back to the agent, delivering it as the return result of the Monitor\_Syscall. The agent can then use the returned data to open the file to be monitored on demand and create a corresponding cache. When a file in a container is accessed, its cache is already present in the kernel, preventing the agent from creating a duplicate cache. 
To monitor such files in real time, this paper adopts an automatic kernel object migration method  proposed in \cite{bib28}. By leveraging VMI technology, the dentry objects of target files are migrated to independent kernel space, where read/write monitoring is performed, ensuring high-performance file monitoring while maintaining system efficiency.

\subsection{Abnormal response}\label{subsec3.4}

Memory events triggered by file access occur in the process of file access system calls, so the monitoring program can prevent unauthorized access according to security rules in real-time. Therefore, the abnormal response module monitors the execution process of the file access system call. When the system call returns in the virtual machine, the abnormal response module sets the system call return value to a negative number. At this time, for the user-mode program, the file access operation fails, and the file handle cannot be obtained to realize further file access. To prevent user-mode programs from guessing the file handle number and accessing files based on the file number, the abnormal response module will clear the relevant memory fields in the operating system kernel. 
Specifically, it targets the dentry and inode objects, By clearing these memory fields and reducing the reference count of these objects, we ensure that any remaining links between the file and the accessing process are severed. When the reference count reaches zero, the operating system automatically reclaims the resources and clears any lingering pointers. This ensures that any resources allocated during the initial file access attempt are safely released, preventing potential resource leaks.

\subsection{ Security analysis and protection}\label{subsec3.5}

The proposed method monitors the target file based on the file metadata cache, so it is necessary to keep the dentry object of the monitored file persisting in the memory of the virtual machine. The operating system periodically cleans up file objects that are no longer accessed (that is, dentry objects with a reference count of 0), so the security protection module needs to prevent the reference count of the target file from dropping to 0. To this end, we propose two methods to protect the monitored dentry objects. Firstly, since all monitored dentry objects are under memory read and write monitoring, changes in the reference count of the target object will trigger monitoring. At this point, the monitoring program checks whether the reference count is set to 0, and if so, resets it to 1 through the VMI technology to ensure that the dentry structure is not cleaned up by the virtual machine. Secondly, the monitored files are all accessed by the security agent/hidden agent, therefore, the security module needs to protect the life cycle of the agent in the virtual machine. To prevent the agent program from being killed by other processes, the security protection module monitors the KILL system call in the virtual machine. When the killed process is an agent process, the security module uses VMI technology to modify the parameters of the system call outside the virtual machine, thereby preventing the agent program from being cleaned up. Since the KILL system call is executed infrequently, the monitor is triggered very infrequently and does not introduce a lot of real-time overhead.

When a container escape attack is performed, the attacker may take full control of the virtual machine, and then attack the agent program running in the virtual machine or the virtual machine kernel. By tampering with the virtual machine kernel, an attacker may replace the virtual machine file system, thus evading monitoring. To this end, the security protection module protects the memory integrity of the virtual machine kernel and the agent program based on the VMI technology. The code segment of the operating system and the agent program should not be tampered with, and the security monitoring module protects the corresponding memory space by setting the corresponding virtual machine memory page to be unwritable. Even if an attacker has control of the virtual machine, there is no way to tamper with the kernel or agents.
In addition, the hypervisor is considered secure, as modern hypervisors provide significantly stronger isolation compared to containers. Although the hypervisor could theoretically be compromised, it offers a robust level of isolation that enhances system security.

Soft and hard links in Linux may affect the security of the system. The soft link of the file is to create a new file path similar to a shortcut to the target file. The parsing of the path will eventually point to the original dentry, so it will not affect the system. The hard link will cause the file to have multiple resolvable paths, and the dentry has a one-to-one relationship with the file path, so a hard link will cause a file to have multiple dentry objects. To solve this problem, when monitoring the existing cached files, the system will find all the related objects according to the doubly linked list between dentry objects and monitor them. Furthermore, the security protection module monitors the LINK system call which is used to create hard links in the virtual machine, so as to capture all new hard links and memory objects to be monitored.

To prevent the monitored object from being migrated to a new memory area by an attacker, the memory monitoring module performs a security check on the reading of the monitored dentry, preventing programs that are not permitted by the security policy from reading the cache of the target file.

\section{System implementation}\label{sec4}
We implement a prototype system based on Intel Virtualization Technology (VT) technology. The virtual machine operating system adopts 64-bit Ubuntu 16.04, and docker is selected as the container engine.

\subsection{Automatically identify kernel structures}\label{subsec4.1}

Semantic gap is a key problem in memory monitoring of virtual machine \cite{bib29}. The monitoring program outside the virtual machine can only obtain binary memory information. In order to map binary memory to high-level meaningful information (such as process name, process number, file name, etc.), the memory distribution knowledge of the virtual machine operating system should be obtained. To this end, we use the Volatility tool \cite{bib30} to automatically obtain the memory distribution knowledge of the operating system. Volatility is a memory analysis tool proposed by Volatility Foundation. It integrates a large number of open source operating system memory distribution configuration files. Based on these configuration files, the relationship between structures and the memory offset of each field in the structure can be obtained automatically. When capturing file access, the monitoring program can obtain information such as CR3 and RSP of the current file access program. Based on these information, the Volatility tool can automatically obtain the program name, namespace, user and other information of the currently accessed program, so as to conduct security analysis.

\subsection{Monitor memory events}\label{subsec4.2}

The memory monitoring module is based on Intel EPT technology, which is a hardware-level memory virtualization technology provided by Intel CPU. Each memory page of the virtual machine has an EPT entry to save its access rights. When the memory access does not match this permission, the VMExit event will be triggered, and the VMExit event will be captured by the VMI module in the virtual machine management layer. The memory monitoring module firstly sets the memory page to be monitored as unreadable or unwritable. When the memory page is accessed by the virtual machine, the monitoring program is triggered and can obtain the accessed memory address. When the memory address is the reference count field of the dentry, the monitor modifies its access rights and makes the CPU single-step, allowing the virtual machine to complete the memory read and write. After single-stepping, the monitor can capture memory changes to analyze whether the operation is a file access operation (reference count increments) or file close operation (reference count decrements). If the file is being accessed, the security of file access is further analyzed based on the security policy.

\subsection{Monitor system calls}\label{subsec4.3}

Monitoring the triggering and returning of system calls can catch specific system calls (such as KILL, LINK, etc.) and prevent abnormal file access. At present, the mainstream 64-bit Linux operating system adopts the fast system call mechanism, and realizes switching between user mode and kernel mode based on SYSCALL and SYSRET instructions. The user-mode program invokes a system call request through the SYSCALL instruction, and calls the corresponding system call according to the system call table after the CPU mode is switched. Based on this mechanism, the system sets a monitoring point at the beginning of the system call to be monitored. By injecting INT3 instructions at the monitoring point, the virtual machine triggers the VMExit event when the monitoring point is executed, so that it can be captured and detected by the monitoring program in the virtual machine management layer. When the system call does not comply the security rules, the parameter is modified to an invalid value based on the VMI technology, so that the system call fails.

The abnormal response of file access occurs during the execution of the system call, so it cannot be blocked at the beginning of the system call. In order to solve this problem, the system blocks at the point where the system call returns. The system captures the event at the virtual machine management layer by replacing the system call return instruction with an INT3 instruction. Due to the context switch mechanism, the first system call return event after a file access event may be the system call return of another process. Therefore, it is necessary to obtain the current CR3 and RSP register values when the system call returns, and compare them with the CR3 and RSP values captured by the file access event, so as to find the correct system call return event and block it.

\section{Evaluation}\label{sec5}

After realizing the prototype system, we test the system in two aspects: effectiveness and performance. The experiment was carried out on a computer with 2.4GHz Intel i5 CPU and 8 GB memory with Intel VT technology. Xen is used to create virtual machines. The host operating system and the virtual machine operating system adopt 64-bit Ubuntu 16.04, and docker is selected for the container engine in the virtual machine.

\subsection{Effectiveness}\label{subsec5.1}

In order to test the effectiveness of the system, we test the protection capabilities of the files in the container and the virtual machine files respectively.

We select the web page file that is easily tampered with in the container and the login executable file as the test object. Website tampering is a common form of network attack, which releases malicious information after penetrating the website server through network attack. Web servers usually store configuration data in specific directories or configuration files, and modifying these files will result in changes to the content and configuration of web pages. In addition, attackers usually replace login related executable files, and implant backdoors in the system to maintain continuous control. Therefore, these files are very important for the security of the system.

We first install Httpd in the container and place the web page file in the www directory, and then configure the web page file and login file in the www directory to be unwritable in the system. After receiving the security policy, the cache creation module establishes the relevant cache and is monitored by the memory monitoring module. The test simulates the scenario where an attacker modifies the web page file and the login file in the container. The result shows that the system can capture the access operation to the target file, and can block the related access in real-time, which verifies the effectiveness of protecting the files in the container.

To verify the ability of the system to protect the files in the virtual machine, we select the runc container escape attack (CVE-2019-5736) as the test object. To defend against this attack, the system configures a security rule to prevent the runc file in the virtual machine from being tampered with. After configuring the security policy, we simulate the attack by accessing the runc file, and the test shows that the system was able to catch and block access to this file.

\subsection{File performance}\label{subsec5.2}

\begin{figure}[h]%
\centering
\includegraphics[width=0.6\textwidth]{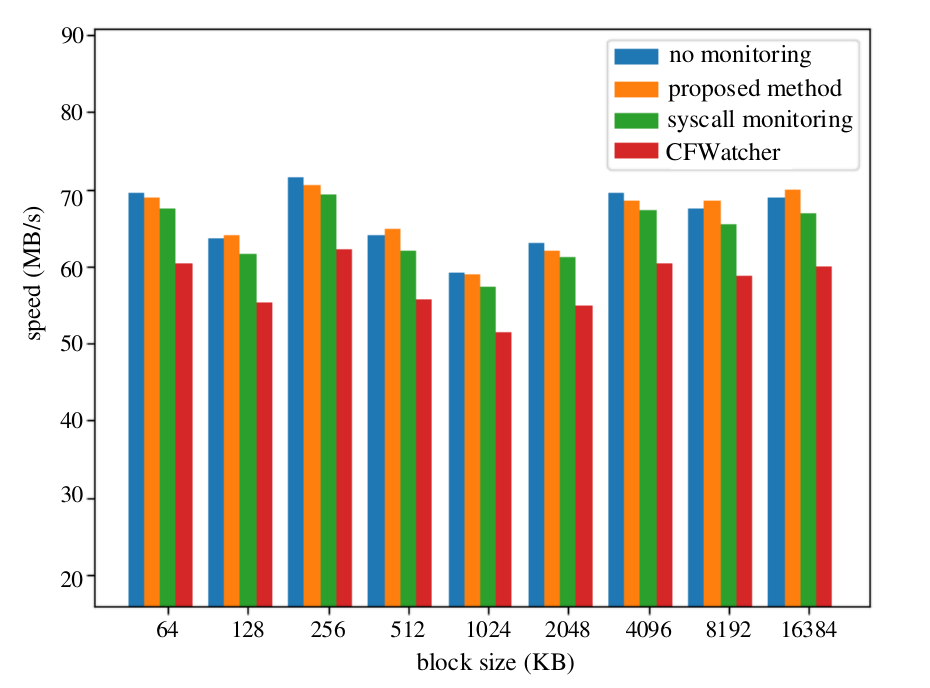}
\caption{Comparison of file read performance of containers under different monitoring methods}\label{fig4}
\end{figure}

\begin{figure}[h]%
\centering
\includegraphics[width=0.6\textwidth]{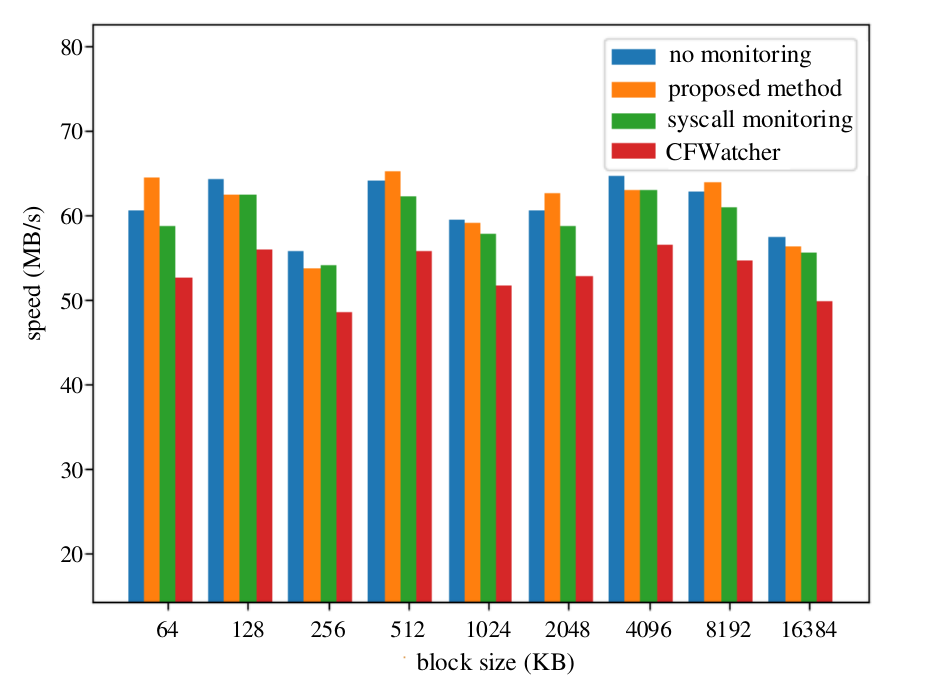}
\caption{Comparison of file writing performance of containers under different monitoring methods}\label{fig5}
\end{figure}

In order to test the impact of the monitoring system on the performance of the container file system, we first use the IOzone tool \cite{bib31} to test the read and write performance of the container file system. First, we pull the IOzone-integrated docker image in Docker Hub to test the read and write performance. The total amount of file read/write is set to 1 GB, and after 100 consecutive tests, the average read and write speeds are shown in Fig.\ref{fig4} and Fig.\ref{fig5} when accessed with different file block sizes. After that, we start the monitoring system and monitored 500 container files. In this case, IOzone and the same test parameters are also used to test the read and write performance of container files.

It can be seen from the experimental results that the difference between the file read/write performance of the container before and after monitoring is within 2\%. 
For instance, under the condition where the block size is 2048KB, the read speed before monitoring is 62.72MB/s, and after applying monitoring, it is 61.69MB/s.

Moreover, in some cases, the read/write performance with monitoring enabled is better than the read/write performance without monitoring enabled,
even if our approach focuses on real-time file access monitoring.
It shows that the performance difference introduced by file monitoring is weaker than the performance fluctuation due to environmental factors. Therefore, the overhead introduced by the proposed file monitoring method is very low. Because the file monitoring method proposed is only related to the monitored target file, and the container's access to other files will not trigger monitoring, no overhead will be introduced. 

The method based on system call interception is widely used in real-time file monitoring. We firstly compare and test this monitoring method. \cite{bib4} injects a kernel module into the virtual machine operating system, which intercepts and analyzes file-related system calls in the virtual machine. In addition, we also compare with real-time file monitoring methods based on VMI. Compared with the system call interception methods based on VMI such as Gemini \cite{bib14}, CFWatcher captures and analyzes the access to target files by monitoring the cache changes of files in the virtual machine, and has a lower load \cite{bib18}. Therefore, we mainly compare with CFWatcher.

The experiment also used IOzone and the same test parameters to test the two monitoring methods 100 times respectively. 
It can be seen from the experimental results that the average performance difference in file read/write speed introduced by the file monitoring inside the virtual machine is about 4\%. CFWatcher introduces a higher monitoring overhead, because the memory monitoring scope of CFWatcher is too large, resulting in a large number of VMExit events being falsely triggered by unrelated file accesses, and the response of VMExit events consumes a lot of time. Compared with these two methods, the proposed file monitoring method based on memory monitoring introduces a lower performance load.

\subsection{Container startup speed}\label{subsec5.3}

The cache creation module will create a cache of monitored files before the container starts, and it will increase the container startup time. In order to test the overhead, we first test the startup time of the container without enabling cache creation, and then test the startup time after enabling the cache creation module. We select the Ubuntu 16.04 image in Docker Hub as the target test container. When the file metadata cache creation is not enabled, the average time for the container to start is 550ms. Since the number of file metadata caches created will affect the running time of the cache creation module, we test the startup time of the container when 100, 200, 300, 400, and 500 file metadata caches are created. We take the average value of 100 times of the test. According to the experimental results, the startup overhead introduced by creating 100-500 file metadata caches are 0.5\%, 0.7\%, 1\%, 1.1\%, and 1.3\% respectively, which are acceptable for the container startup process. 
For example, when 500 file metadata caches are created, the container startup time is approximately 557ms.
The average creation time of each file metadata cache of CFWatcher is 2ms, which is longer than the proposed creation method. This is because CFWatcher needs to wait for the right injection time and needs to intervene the running of the target virtual machine based on the VMI, both of which can consume a lot of time.

\begin{figure}[h]%
\centering
\includegraphics[width=0.7\textwidth]{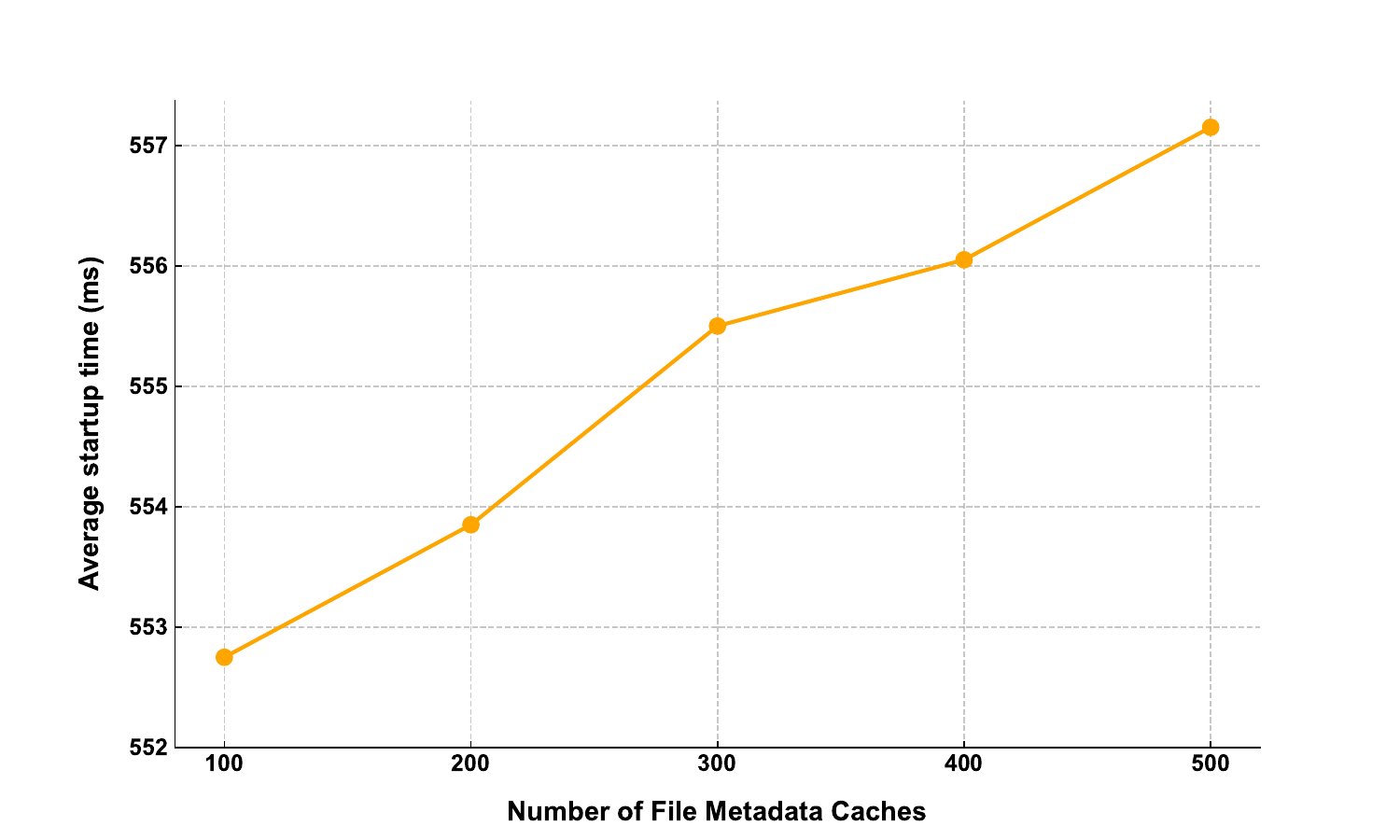}
\caption{Impact of file metadata cache count on average startup time}\label{fig6}
\end{figure}

\begin{figure}[h]%
\centering
\includegraphics[width=0.7\textwidth]{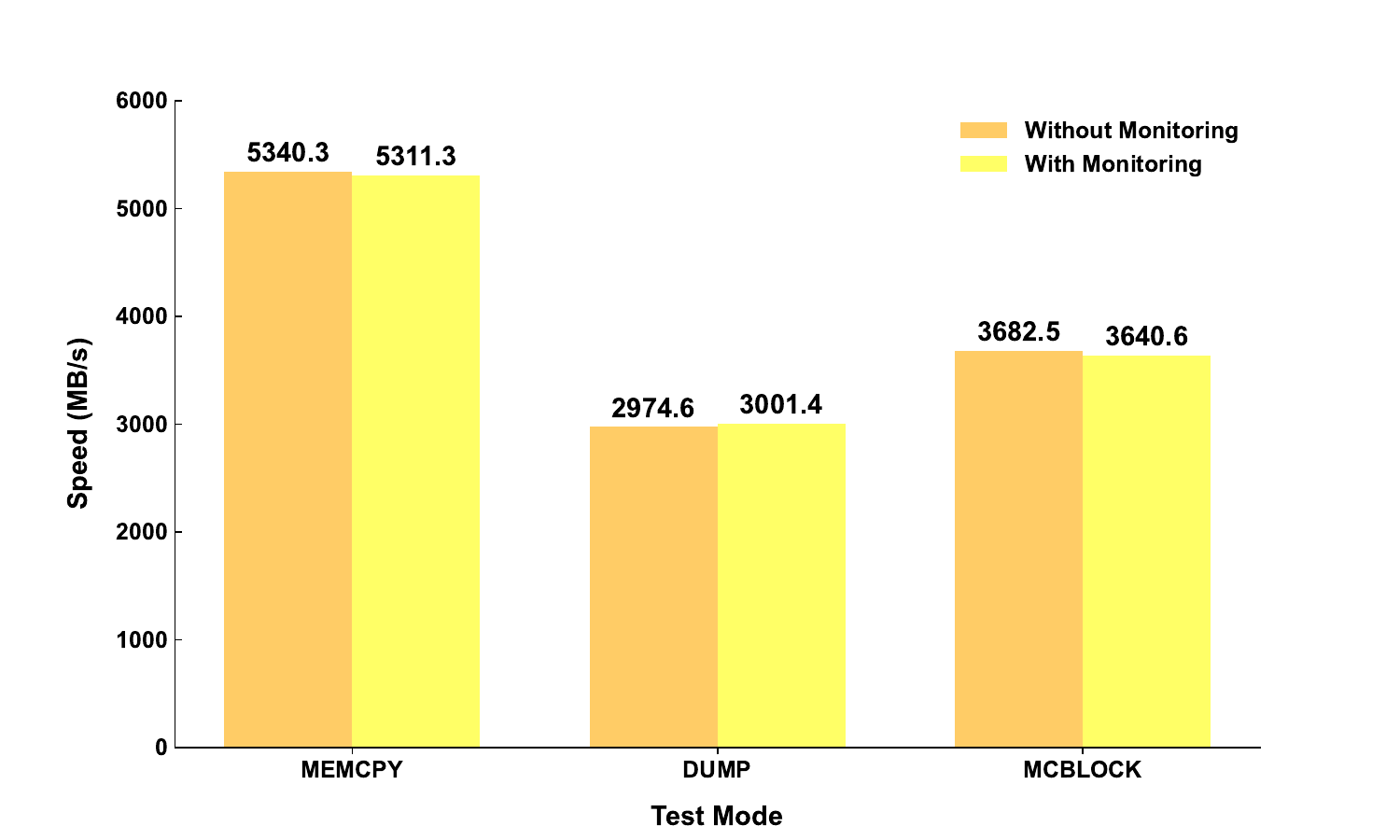}
\caption{Effect of monitoring on memory performance in various test modes}\label{fig7}
\end{figure}

\subsection{Memory performance }\label{subsec5.4}

The memory monitoring module will monitor part of the memory of the target virtual machine, so we test the impact of the monitoring system on the memory performance of the virtual machine. In the experiment, the memory test tool mbw \cite{bib32} is used to test the memory performance of the container in the virtual machine. 
This tool simulates memory read and write operations by copying arrays between different memory blocks, thus measuring memory performance. Specifically, mbw measures the transfer speeds of memory during consecutive copy operations across three testing modes: MEMCPY (memory copy), DUMP (string copy), and MCBLOCK (memory block copy).
The amount of memory read/write tested is 128MB and the average speeds of 100 copies without monitoring are 5,340.3MB/s (MEMCPY), 2974.6MB/s (DUMP) and 3682.5 MB/s (MCBLOCK); when the monitoring is running, the average speeds are 5,311.3MB/s (MEMCPY), 3001.4MB/s (DUMP), and 3640.6 MB/s (MCBLOCK), respectively. 
It can be seen from the results that the memory performance difference before and after monitoring is around 1\%. When the test mode is MCBLOCK, the speed is even faster than the pre-monitoring speed. Therefore, the impact of monitoring on container memory performance is weaker than that of environmental interference factors.

The container files monitor based on file caching, so it will occupy additional memory areas. Usually the directory cache size of a file is 192 Bytes, and the file metadata cache created by the cache creation module is stored continuously, so the memory required to monitor 500 files occupies about 24 memory pages, with a total size of 96KB. Compared to the total memory of the container, the memory overhead introduced is acceptable.

\section{Discussion}\label{sec7}

In our current work, we focus on monitoring critical files within containers to ensure high efficiency and minimize performance overhead, providing robust protection against unauthorized access and tampering. However, we recognize that container security extends beyond file integrity. Future work will expand monitoring to include system calls, network activities, and container configurations, enabling us to detect and prevent threats like privilege escalation, unauthorized network access, and misconfigurations. By broadening our monitoring scope, we aim to enhance container security and resilience without introducing significant performance overhead. We will optimize this expansion through adaptive monitoring based on threat levels and resource availability. In addition, we will expand our monitoring capabilities using machine learning-based anomaly detection to identify subtle patterns of potential threats. Additionally, we will enhance response strategies with dynamic, adaptive mechanisms, enabling tailored interventions such as adjusting security settings or isolating compromised containers to minimize impact and maintain operational continuity.

Another limitation is that our monitoring system triggers processes for both read and write operations, ensuring all file interactions are captured for security analysis. However, this design introduces performance overhead, especially in read-heavy environments where frequent read operations may incur unnecessary monitoring costs. Since read accesses are generally less critical than writes, this approach may not be optimal. We acknowledge this limitation and plan to explore selective monitoring strategies, such as prioritizing write operations or adjusting based on file type and data sensitivity. This would reduce overhead in read-dominant scenarios while preserving security.

Finally, usability is a recognized limitation of our approach. The current approach requires modifications to the virtual machine kernel and container image to support extra system calls and configuration policies. While this ensures robust security, it adds complexity to the system and reduces ease of use. Future research could explore ways to simplify the integration process without compromising on security.This approach also introduces a limitation when a container attempts to acquire exclusive locks on files that are already opened by the security agent. Since these files remain open, the container's attempts to acquire exclusive locks may fail, which could affect operations that require exclusive access to certain files.

\section{Conclusion}\label{sec8}

A real-time container file monitoring method is proposed based on VMI. Due to the strong isolation ability of the hypervisor, the monitoring program can resist the security threats brought by container escape attacks. After analyzing the container file system, the cache object of the target file is monitored, so that the monitoring program is only triggered by the access operation of the target file, which reduces the performance overhead introduced by real-time file monitoring. In addition, an automatic file metadata cache creation approach is proposed based on the cooperation between virtual machine and hypervisor, so that file monitoring can meet the container startup speed requirements of container scenarios. 
Experimental results that the method can effectively monitor critical container files and prevent them from being maliciously tampered with. The difference in file read/write performance of the container before and after monitoring is minimal, and the introduced monitoring overhead is acceptable.

\section*{Declarations}

\subsection*{Funding}

This work was supported by the National Natural Science Foundation of China under Grants No. 62302122 and No. 62172123, the Natural Science Foundation of Heilongjiang Province of China under Grants No. LH2023F017, and CCF-Huawei Populus Grove Fund under Grants No. CCF-HuaweiSY202411.

\subsection*{Conflict of interest/Competing interests}

The authors declare that they have no conflicts of interest to report regarding the present study.

\subsection*{Ethics approval}

The study was approved by the Harbin Institute of Technology.

\subsection*{Consent to participate}

Informed consent was obtained from all participants included in the study.

\subsection*{Data availability}

The datasets generated during and/or analysed during the current study are available from the corresponding author on reasonable request.



\subsection*{Authors' contributions}

Kai Tan and Dongyang Zhan conceived and designed the study. Lin Ye, Hongli Zhang and Binxing Fang collected the data. Kai Tan and Dongyang Zhan analyzed and interpreted the data. Kai Tan drafted the manuscript. Zhihong Tian helped revise the paper.

All authors critically reviewed the manuscript and approved the final version for publication.

\noindent

\bigskip
\begin{flushleft}%
Editorial Policies for:

\bigskip\noindent
Springer journals and proceedings: \url{https://www.springer.com/gp/editorial-policies}

\bigskip\noindent
Nature Portfolio journals: \url{https://www.nature.com/nature-research/editorial-policies}

\bigskip\noindent
\textit{Scientific Reports}: \url{https://www.nature.com/srep/journal-policies/editorial-policies}

\bigskip\noindent
BMC journals: \url{https://www.biomedcentral.com/getpublished/editorial-policies}
\end{flushleft}

\bibliography{sn-bibliography}

\end{document}